%% file: elsarticle-template-harv.tex
\documentclass[final,3p,times,twocolumn]{elsarticle} 
\newcommand{\TableFontSize}{\small}

\usepackage{amssymb}
\usepackage{amsmath}
\usepackage{siunitx}
\usepackage{mhchem}
\usepackage{subcaption}
\usepackage{xfrac} 
\usepackage{framed}
\usepackage{multicol}
\usepackage{nomencl}
\usepackage[outdir=./]{epstopdf}

\makenomenclature

\newif\ifnomentry

\ExplSyntaxOn 
\NewExpandableDocumentCommand{\stringcase}{mm}
 {
  \str_case:nn { #1 } { #2 }
 }
\ExplSyntaxOff

\journal{arXiv}

\makenomenclature
\renewcommand*\nompreamble{\begin{multicols}{2}}
\renewcommand*\nompostamble{\end{multicols}}

\begin{document}
\begin{frontmatter}

\title{Thin-film boundary-layer diffusion of non-equilibrium flow to kinetically limited reactive surfaces via Damk\"ohler thermochemistry tables}

\author{Jeffrey D. Engerer and Lincoln N. Collins} 

\affiliation{organization={Engineering Sciences, Sandia National Laboratories},
            addressline={1515 Eubank SE}, 
            city={Albuquerque, NM},
            postcode={87123}, 
            state={NM},
            country={USA}}

\begin{abstract}
Traditional ablation thermochemistry tables for atmospheric entry are derived from boundary-layer element diffusion assuming homogeneous and heterogeneous thermochemical equilibrium at the vehicle surface. 
Prior techniques for finite-rate surface reactions predominantly embed specific heterogeneous reaction models within the homogeneous equilibrium solution procedures and tables.
This paper disseminates a boundary-layer integral solution for wall-gas free oxygen coupled to finite-rate surface kinetics.
Solutions are pre-tabulated along normalized kinetics variables without direct integration into an equilibrium thermochemistry solver. 
This technique allows greater flexibility in presumed kinetics rates and wall-gas conditions in simple air-carbon systems, but the extensibility to state-of-the-art simulations and complex materials remains uncertain. 
A derivation and preliminary results are presented to the encourage further development.
\end{abstract}




\end{frontmatter}

\begin{table*}[!t]
\TableFontSize
   \begin{framed}

\nomenclature[C]{$g_m$}{mass conductance}
\nomenclature[C]{$\mathit{St}_m$}{mass-transfer Stanton number}
\nomenclature[G]{$\rho$}{Density}
\nomenclature[C$y$]{$\tilde{y}$}{elemental mass fraction}
\nomenclature[C]{$y$}{species mass fraction}
\nomenclature[C]{$u$}{velocity}
\nomenclature[C]{$k$}{kinetic rate}
\nomenclature[C]{$n,m$}{reaction order}
\nomenclature[C]{$m_c''$}{ablative mass flux}
\nomenclature[C]{$M$}{molar mass}
\nomenclature[C$D$]{$\overline{Da}$}{oxygen-consumption Damk\"ohler number}
\nomenclature[C]{$B_c'$}{non-dim. ablation rate}
\nomenclature[C]{$B_{C,s}'$}{non-dim. direct carbon flux}
\nomenclature[C$j$]{$\tilde{j}$}{elemental diffusive flux}
\nomenclature[G]{$\Delta_{\mathrm{C:O}}$}{wall product ratio}
\nomenclature[G]{$\Gamma_O$}{transition function, oxygen}
\nomenclature[G]{$\Gamma_C$}{transition function, carbon}
\nomenclature[G]{$\gamma$}{reaction multiplier}
\nomenclature[G]{$\sigma_\mathbb{N}$}{model error term}
\nomenclature[C]{$T$}{temperature}
\nomenclature[C]{$P$}{pressure}
\nomenclature[C]{$K_p$}{equilibrium constant}
\nomenclature[G]{$\varepsilon$}{reaction efficiency}
\nomenclature[C]{$R$}{ideal gas constant}
\nomenclature[C]{$r$}{stagnation-point radius}

\nomenclature[Y]{$w$}{wall gas}
\nomenclature[Y]{$e$}{edge gas}
\nomenclature[Y]{$\mathrm{O}_{diss}$}{dissociated oxygen}
\nomenclature[Y]{$\mathrm{O}_x$}{free oxygen}
\nomenclature[Y]{$\mathrm{C}_x$}{free carbon}
\nomenclature[Y]{$c$}{ablating surface}
\printnomenclature     
   \end{framed}
\end{table*}


\section{Introduction}
\label{sec:intro}
\input{intro.tex}

\section{Derivation}
\label{sec:derivation}

\input{derivation.tex}

\section{Results}
\label{sec:demonstration}

\input{demonstration.tex}

\input{conclusion.tex}

\section*{Acknowledgement}
The authors thank John-Paul Heinzen and Rachel Hays for discussions and collaborations that improved the quality of this work.
Supported by the LDRD Program at Sandia National Laboratories. Sandia is managed and operated by NTESS under DOE NNSA contract DE-NA0003525.


\bibliographystyle{elsarticle-harv} 
\bibliography{library}

\end{document}

%% file: intro.tex
Ablative atmospheric entry produces chemical reactions at the vehicle surface coupled to boundary-layer heat and mass transfer. 
Traditional analysis techniques are classified by the resolution of the boundary-layer physics: differential methods resolving boundary-layer species and reactions, 
and integral methods characterizing transport via the potential difference and boundary-layer conductance. 
The developments herein target the latter class of methods, but could be extended to physics coupling in the former class. 

Restricting the present analysis to mass-transfer, the mass-conductance of the ablative boundary layer, $g_m$, is traditionally defined via the mass-transfer Stanton number, $\mathit{St}_m$:
\begin{equation}
g_m = \rho_e u_e \mathit{St}_m
\end{equation}
where $\rho_e$ and $u_e$ are the density and velocity of the gas at the outer edge of the boundary layer.

Under the assumption of equal species diffusion coefficients, boundary-layer element transport (e.g., C, N, O) is effectively a linear superposition of species carrying those elements and thereby remains insensitive to the particular chemical species and reactions present throughout the boundary-layer. 
The driving potential for element transport is the mass fraction, $\tilde{y}_k$, of a given element,  $k$, summed across all species:
\begin{equation}
	\tilde{y}_k = \textstyle \sum_{j} \tfrac{n_{k,j} M_k}{M_j} y_j 
 \label{eqn:species_mass_fraction}
\end{equation}
where $y_j$ and $M_j$ are the mass fraction and molar mass of species $j$, and $n_{k,j}$ and $M_k$ are the number and molar mass of element $k$ within species $j$. 
The resulting elemental diffusive flux ($\tilde{j}_{k}$) is:
	\begin{equation}
	\tilde{j}_{k} = \rho_e u_e \mathit{St}_m (\tilde{y}_{k,w} - \tilde{y}_{k,e})
	\label{eqn:element_diffusion}
\end{equation}
where subscripts are $w$ for wall gas and $e$ for edge gas.

The ablative mass-flux, $m''_c$, is normalized by the mass conductance, yielding the non-dimensional ablation rate, $B'_c=m''_C/g_m$. 
For non-decomposing ablators, the coupling of the boundary-layer diffusion to surface mass flux yields the elemental balance:
\begin{equation}
 	\tilde{y}_{k,w} = \frac{\tilde{y}_{k,e} + B'_c\tilde{y}_{k,c}}{1+B'_c} 
 	\label{eqn:element_balance}
 	\end{equation}
where $\tilde{y}_{k,c}$ represents elemental composition of the ablating surface. 
Provided the thermodynamic state, the wall-gas species and enthalpy are determined via thermochemical models, typically assuming homogeneous chemical equilibrium. 
Therefore, if surface ablation and wall-gas thermochemistry depends on temperature, $T$, pressure, $P$, and a limited set of additional variables, solutions can be pre-tabulated. 

Traditional thermochemistry tables assume homogeneous thermochemical equilibrium within the wall-gas and heterogeneous equilibrium with the ablating surface, rendering $B'_c$ a function of $T$ and $P$ alone. 
These assumptions yield the well-established diffusion-limited result: oxygen-limited ablation for CO$_2$ and CO product augmented by carbon sublimation at high temperature. 
The equilibrium approach provides rapid, stable solutions but neglects the kinetic limitations of surface reactions. 
Past boundary-layer integral solutions for finite-rate gas-surface interactions embed kinetics models within numerical methods computing homogeneous thermochemical equilibrium \cite{Milos1997}. 
This technique has been fully deployed with pretabulated and table-free approaches \cite{Milos2013, Milos2012}, but has tied finite-rate ablation to intricate thermochemistry solvers. 

Empirical approximations of finite-rate graphite ablation compute the overall oxidation rate via a resistors-in-series analogy. 
In this model, the resistances of oxygen diffusion and surface oxidation are intermediated by the oxygen partial pressure in the gas immediately external to the ablating surface (i.e., wall gas).
The ratio of reaction-limited ($m''_R$) to diffusion-limited ($m''_D$) ablation rates is computed and applied to one of several proposed transition functions \cite{Frank-Kamanetskii1969,Scala1965,Perini1971} while assuming a CO ablation product.
This approach avoids wall-gas thermochemistry solutions with approximations favorable to air-carbon systems. 

This empirical approximation has been demonstrated for first-order and half-order kinetics models \cite{Welsh1963,Metzger1967,Scala1965,Perini1971}. Metzger \cite{Metzger1967} found excellent agreement to graphite ablation experiments using half-order kinetics, although the analysis of Perini \cite{Perini1971} demonstrates suppressed ablation rates in higher-pressure facilities ($P$$\gtrsim$\SI{1}{atm}).
This high-pressure trend appears consistent with the double-plateau theory established by Welsh \cite{Welsh1963}, wherein ablation rate at intermediate temperature (\SIrange{1500}{2500}{K}) is reduced by CO$_2$ ablation product.
However, the empirical ablation-ratio approximation cannot predict this variance in ablation product while neglecting the wall-gas thermochemistry. 

This paper extends the ablation-ratio methodology \cite{Scala1965,Perini1971,Metzger1967} by integrating wall-gas thermochemistry solutions. 
This approach arrives at two non-dimensional kinetics variables enabling pre-tabulated thermochemistry, namely: the oxygen-consumption Damk\"ohler number ($\overline{Da}$) and an arbitrary source of combustible carbon ($B'_{C,s}$). 
The proposed Damk\"ohler number appears broadly applicable for empirical and computational purposes.
Relative to current techniques \cite{Milos1997,Milos2012,Milos2013}, the proposed approach improves the flexibility of kinetics models and isolates the ablation-rate calculation from the thermochemistry solver.

%% file: derivation.tex
Here, the resistors-in-series analogy of prior empirical works is rederived from the traditional governing equations of ablation thermochemistry.
Oxygen boundary-layer transport is computed from element diffusion, kinetics are non-dimensionalized for table generation, and the wall-gas thermochemical state is resolved. 

\subsection{Diffusion of Free Carbon and Oxygen}

First, the derivation mathematically isolates boundary-layer diffusion of free carbon (C$_x$) and free oxygen (O$_x$) from the ablation product (CO, CO$_2$). 
Kays \cite{Kays_ch16} demonstrates a carefully selected linear combination of elements simplifies the analysis of chemically reactive boundary-layers under the equal species-diffusion assumption.
Therefore, the pseudo-element $\xi$ is defined as a combination of carbon and oxygen element:
\begin{equation}
	\tilde{y}_\xi = \tilde{y}_\mathrm{C} - \Delta_{\mathrm{C:O}} \tilde{y}_\mathrm{O}
\end{equation}
where $\Delta_{\mathrm{C:O}}$ is the carbon-to-oxygen mass ratio of product gases (CO, CO$_2$) at the wall, explicitly: \begin{equation}
	\Delta_{\mathrm{C:O}} = \frac{ \frac{12}{28} y_{\mathrm{CO},w} + \frac{12}{44} y_{\mathrm{CO}_2,w}}
 {\frac{16}{28} y_{\mathrm{CO},w} + \frac{32}{44} y_{\mathrm{CO}_2,w}}
 \label{eqn:Delta}
\end{equation}
The driving potential for boundary-layer diffusion of this pseudo-element is determined from the mass fractions of carbon and oxygen element (Eq.~\ref{eqn:species_mass_fraction}) within $\xi$: 
\begin{equation}
	\tilde{y}_\xi = \textstyle \sum_{j} \tfrac{n_{C,j} M_C}{M_j} y_j - \Delta_{C:O} \textstyle \sum_{j} \tfrac{n_{O,j} M_O}{M_j} y_j 
 \label{eqn:pseudoelement_mass_fraction}
\end{equation}
Assuming the atmosphere contains negligible carbon, the edge-gas
($\tilde{y}_{\xi,e}$) and wall-gas ($\tilde{y}_{\xi,w}$) mass fractions of $\xi$ are:
\begin{equation}
\tilde{y}_{\xi,e} =  -\Delta_{\mathrm{C:O}}\tilde{y}_{\mathrm{O},e}
\end{equation}
\begin{equation}
\tilde{y}_{\xi,w} = \Sigma y_{\mathrm{C}_x,w}-\Delta_{\mathrm{C:O}}\Sigma y_{\mathrm{O}_x,w} +  \textstyle \sum_{j\in \mathbb{N}} \left( \frac{n_\mathrm{C} M_\mathrm{C}}{M_j}  - \Delta_{\mathrm{C:O}}  \frac{n_\mathrm{O} M_\mathrm{O}}{M_j} \right)y_{j,w}
\end{equation}
where notably the contributions of the reaction product (CO and CO$_2$) in the wall-gas are nullified. 
Species set $\mathbb{N}$ contains oxygen and carbon bonded to nitrogen (e.g., CN, NO), modulating ablation with the summation henceforth referred to as $\sigma_\mathbb{N}$.
 
The linear superposition principle yielding the element diffusion equation (Eq.~\ref{eqn:element_diffusion}) remains valid for the defined linear combination of elements, $\xi$.
The mass flux of $\xi$ produced by the ablating surface is that of the carbon element alone ($m''_\xi=m''_c$).
The utility of $\xi$ concludes by applying the conservation of $\xi$ at the ablating surface, yielding the non-dimensionalized ablation rate: 
	\begin{equation}
	    B'_c =  \Delta_{\mathrm{C:O}} (\tilde{y}_{\mathrm{O},e} - \textstyle \Sigma y_{\mathrm{O}_x,w} ) +  \Sigma y_{\mathrm{C}_x,w} + \sigma_\mathbb{N}
     \label{eqn:Bprime_diffusion}
\end{equation}
This solution reveals a simple principle: carbon ablation rate is the summation of (1) carbon consumed by the diffusive flux of oxygen, (2) carbon-vapor diffusion, and (3) secondary effects via nitrogen reactions with carbon and oxygen. 
The driving potentials isolate free oxygen and free carbon, enabling the surface-kinetics implementation.

Free oxygen and free carbon are highly reactive and unlikely to coexist in significant quantities. 
Defining surface mass flux as the summation of oxidative, $m''_{\mathrm{C},oxy}$, and direct (e.g., sublimation, spallation), $m''_{\mathrm{C},s}$ sources, the direct-carbon source must overwhelm the diffusion of oxygen to attain a carbon-rich state.
Thereby, ablation-rate  solutions are divided into carbon-rich and oxygen-rich regimes: 
\begin{equation}
B'_c = 
\begin{cases}
    \Delta_{\mathrm{C:O}} (\tilde{y}_{\mathrm{O},e} - \Sigma y_{\mathrm{O}_x,w} ) + 
    \sigma_{\mathbb{N}}
    &
     \text{if $B'_{\mathrm{C},s}\leq\Delta_{\mathrm{C:O}}\tilde{y}_{\mathrm{O},e} + \sigma_{\mathbb{N}}$
    }
    \\
    \Delta_{\mathrm{C:O}} \tilde{y}_{\mathrm{O},e} + \Sigma y_{\mathrm{C}_x,w}+ 
    \sigma_{\mathbb{N}}
    &
     \text{if $B'_{\mathrm{C},s}>\Delta_{\mathrm{C:O}}\tilde{y}_{\mathrm{O},e} + \sigma_{\mathbb{N}}$
    }  
\end{cases}
\label{eqn:Bprime_regimes}
\end{equation}
where $B'_{\mathrm{C},s}=m''_{\mathrm{C},s}/g_m$ is the direct carbon source normalized by mass conductance.
The expression for the oxygen-rich regime is foundational for the present derivation of kinetically limited thermochemistry solutions, where traditional tables predict only diffusion-limited ablation (CO and CO$_2$ plateaus). 
Meanwhile, the carbon-rich regime describes the sublimation region of traditional tables, but is not of present focus.

\subsection{Coupling to Surface Kinetics}

Second, the derivation defines non-dimensionalized surface kinetics terms compatible with pretabulation. 
Direct carbon flux, $B'_{\mathrm{C},s}$ remains entirely arbitrary. Oxidation kinetics are coupled to diffusion via the free-oxygen content at the wall. 
The surface oxidation rate for arbitrary order kinetics is:
\begin{equation}
	m''_{oxy} =  k_{\mathrm{O}} \left(\rho_w y_{\mathrm{O},w} \right)^m + k_{\mathrm{O}_2} \left( \rho_w y_{\mathrm{O}_2,w} \right)^n = \bar{k}_{\mathrm{O}_x} \left(\rho_w \Sigma{y}_{\mathrm{O}_x,w} \right)^n
	\label{eqn:nth_order_kinetics}
\end{equation}
where $\bar{k}_{\mathrm{O}_x}$ is the averaged oxygen consumption kinetics weighted by the relative mass concentrations of O and O$_2$:
\begin{equation}
\bar{k}_{\mathrm{O}_x} =   k_\mathrm{O} \left( \rho_w \Sigma y_{\mathrm{O}_x,w}\right)^{m-n}
\hat{y}_{\mathrm{O}_{diss}}^m
+  k_{\mathrm{O}_2} 
\left(1-\hat{y}_{\mathrm{O}_{diss}}\right)^n
\label{eqn:averaged_oxidation}
\end{equation}
where $\hat{y}_{\mathrm{O}_{diss}}$ is the ratio of dissociated oxygen to free oxygen. 

Carbon consumption varies from oxygen kinetics by the carbon-to-oxygen ratio in the heterogeneous reaction product: $\Delta_{\mathrm{C:O},r}$. 
For example, the surface might produce CO ($\Delta_{\mathrm{C:O},r}=0.75$), while the wall-gas is dominated by CO$_2$ ($\Delta_{\mathrm{C:O}}=0.375$).
Normalizing these kinetics terms by mass conductance yields:
\begin{equation}
B'_c = \Delta_{\mathrm{C:O},r} \frac{\bar{k}_{\mathrm{O_x}}}{g_m} \left( \Sigma y_{\mathrm{O}_x,w}\right)^n +
 B'_{\mathrm{C},s}
    \label{eqn:Bprime_kinetics}
\end{equation}

Defining $\gamma=\Delta_{\mathrm{C:O},r}/\Delta_{\mathrm{C:O}}$, boundary-layer diffusion (Eq.~\ref{eqn:Bprime_regimes}) is coupled to kinetics (Eq.~\ref{eqn:Bprime_kinetics}), yielding:
\begin{equation}
\frac{\Sigma y_{\mathrm{O}_x,w}}{\tilde{y}_{O,e}}+ \gamma \overline{Da} \left( \frac{\Sigma y_{\mathrm{O}_x,w}}{\tilde{y}_{O,e}}\right)^n - \left( 1-
 \frac{B'_{\mathrm{C},s}}{\Delta_{\mathrm{C:O}} \tilde{y}_{O,e}}\right)
=
 \frac{\sigma_\mathbb{N}}{\Delta_{\mathrm{C:O}} \tilde{y}_{O,e}}
    \label{eqn:interfacecoupling}
\end{equation}
where the oxygen-consumption Damk\"ohler number is:
\begin{equation}
\overline{Da} = \left({\rho_w}\tilde{y}_{\mathrm{O},e}\right)^{(n-1)} \tfrac{  \rho_w \bar{k}_{\mathrm{O}_x}}{ \rho_e u_e \mathit{St}_m }
\label{eqn:Damkohler}
\end{equation}
This Damk\"ohler number replaces the ablation-rate ratio ($m''_R/m''_D$) from prior works \cite{Scala1962,Perini1971,Metzger1967}. 
Notably, the ablation-ratio method assumed a particular wall-gas chemistry state: constant $\Delta_{C:O}$ and wall-gas molecular weight matching undissociated air: $M_w\approx M_{ref}$ \cite{Perini1971}. Therefore, the oxygen-consumption Damk\"ohler number compares to prior works by the relation:
\begin{equation}
    \gamma \overline{Da} = \left(\frac{M_{w}}{M_{ref}}\right)^n\frac{m''_R}{m''_D}
    \label{eqn:Da_compare}
\end{equation}

\subsection{Ablation Rate Solutions}
Neglecting the contribution of $\sigma_\mathbb{N}$, first-order kinetics applied to Eq.~\ref{eqn:interfacecoupling} yields:
\begin{equation}
    B'_c = \Delta_{\mathrm{C:O}} \Gamma_\mathrm{O} \tilde{y}_{\mathrm{O},e} + \Gamma_\mathrm{C} B'_{\mathrm{C},s}
    \label{eqn:Bprime}
\end{equation}
\begin{align}
    \Gamma_\mathrm{O} = \frac{1}{1+(\gamma \overline{Da})^{-1}} \quad; \quad     
    \Gamma_\mathrm{C} = \frac{1}{1+\gamma \overline{Da}}
    \label{eqn:Gamma_firstOrder}
\end{align}
where the functions $\Gamma_\mathrm{O}$ and $\Gamma_\mathrm{C}$ quantify the transition of oxidative and direct-carbon mechanisms from kinetically limited to diffusion-limited regimes.

Scala \cite{Scala1965} and Welsh \cite{Welsh1963} demonstrate improved performance with half-order kinetics. 
The half-order solution to Eq.~\ref{eqn:interfacecoupling} yields the transition functions:
\begin{align}
     \Gamma_\mathrm{O} = &
  \frac{1}
 {\frac{1}{2}+ \sqrt{\frac{1}{4}+ 
 \left(1-\frac{B'_{\mathrm{C},s}}{\Delta_{\mathrm{C:O}}\tilde{y}_{\mathrm{O},e}}\right)\left(\gamma \overline{Da}\right)^{-2}}} 
 \label{eqn:Gamma_halfOrder} \nonumber
\\
\Gamma_\mathrm{C} =  &
\frac{
  \left(1-\frac{B'_{\mathrm{C},s}}{\Delta_{\mathrm{C:O}}\tilde{y}_{\mathrm{O},e}}\right) 
  \left(\gamma \overline{Da}\right)^{-2}
   } 
     { \left( 
 		\frac{1}{2} +
 		\sqrt{\frac{1}{4} + \left(1-\frac{B'_{\mathrm{C},s}}{\Delta_{\mathrm{C:O}}\tilde{y}_{\mathrm{O},e}}\right)\left(\gamma \overline{Da}\right)^{-2}} 
         \right)^2}
\end{align}
For the case where $B'_{\mathrm{C},s}=0$, 
$\Gamma_\mathrm{O}$ matches Perini \cite{Perini1971}, and is often further simplified to the half-order resistors-in-series model \cite{Scala1965,Metzger1967}:  $\Gamma= [1+(m''_D/m''_R)^2]^{-\sfrac{1}{2}}$. 

\subsection{Table Generation and Look-up}

As with traditional tables, the free-stream oxygen elemental content is a predefined constant (e.g., $\tilde{y}_{\mathrm{O},e}=0.233$ for air).
Therefore, solution to Eq.~\ref{eqn:Bprime} requires two kinetics variables ($\gamma \overline{Da}$, $B'_{\mathrm{C},s}$) and the thermodynamic state ($T,P$).  
A generalized solution procedure for thermochemistry table generation is:
\begin{enumerate}
    \item Generate table array [$T, P, \gamma\overline{Da}, B'_{\mathrm{C},s}$].
    \item Estimate $M_w$ and $\Delta_{\mathrm{C:O}}$.
    \item Calculate Gamma functions via Eq.~\ref{eqn:Gamma_firstOrder} or~\ref{eqn:Gamma_halfOrder}. 
    \item Compute $B'_c$ via Eq.~\ref{eqn:Bprime}.
    \item Solve elemental balance (Eq.~\ref{eqn:element_balance}).
    \item Convergence on Steps 2-5, as required.
\end{enumerate}
Arbitrary assumptions for non-equilibrium gas chemistry are acceptable, such as tables that force CO ablation product (e.g., $\Delta_{\mathrm{C:O}}=0.75$). 

At table lookup, $B'_{\mathrm{C},s}$ can include any mechanism delivering carbon to the surface expected to react entirely with the free oxygen at the wall. 
For example, carbon spall from an arbitrary model form can be combusted entirely or partially.
Oxidation rate, $\bar{k}_{\mathrm{O}_x}$, can be arbitrarily defined, provided the reaction order is consistent with the lookup table.
If either kinetics model requires the wall gas state, (e.g., $\hat{y}_{\mathrm{O}_{diss}}$, $\Delta_{\mathrm{C:O}}$, $y_j$), additional table outputs are tabulated.

%% file: demonstration.tex
A prototypical workflow is presented here to demonstrate requirements, flexibility, and predictions.
The kinetics models of prior works using the ablation-ratio technique are prioritized for direct comparison to the prior methodology.

\subsection{Non-Equilibrium Thermochemistry Tables}
\label{sec:tablegen} 
Thermochemistry tables are generated for homogeneous equilibrium and for a non-equilibrium case suppressing CO$_2$.
Extending the approach of Welsh \cite{Welsh1963}, equilibrium is selectively applied with the following reaction set:
\begin{align}
   \ce{O_2(g) <=> 2O(g)} & 
   \hspace{10pt} 
   K_{p,1}=\tfrac{P^2_\mathrm{O}}{P_{\mathrm{O}_2}} 
   \label{eqn:Kp_1}
   \\
   \ce{2\mathrm{CO}_2(g) <=> 2CO(g) +O_2(g)} & 
   \hspace{10pt} 
   K_{p,2}=\tfrac{P_{\mathrm{CO}}^2 P_{\mathrm{O}_2}}{P^2_{\mathrm{CO}_2}}
   \label{eqn:Kp_2}
   \\
   \ce{CO_2(g) <=> CO(g) +O(g)} & 
   \hspace{10pt} 
   K_{p,3}=\tfrac{P_{\mathrm{CO}}P_{\mathrm{O}}}{P_{\mathrm{CO}_2}}
   \label{eqn:Kp_3}
\end{align}    
where $N_2$ dissociation and reactions are ignored ($\sigma_\mathcal{N}=0$). 
The resulting equilibrium constants of Eqs.~\ref{eqn:Kp_1}--\ref{eqn:Kp_3} predict:
\begin{equation}
    \hat{y}_{\mathrm{O},diss} = \left(1+2\sqrt{P_{\mathrm{O}_2}/K_{p,1}} \right)^{-1}
    \label{eqn:Odiss_Kp}
\end{equation}
\begin{equation}
    \Delta_{\mathrm{C:O}} = \frac{\sqrt{P_{\mathrm{O}_2}}+ \sqrt{ K_{p,1}}
    \tfrac{K_{p,2}}{K_{p,3}}}
    {2\sqrt{P_{\mathrm{O}_2}}+ \sqrt{ K_{p,1}}
    \tfrac{K_{p,2}}{K_{p,3}}}
    \frac{M_\mathrm{C}}{M_\mathrm{O}}
    \label{eqn:Delta_Kp}
\end{equation}
The full equilibrium system uses Eqs.~\ref{eqn:Odiss_Kp} \&~\ref{eqn:Delta_Kp}; the non-equilibrium case equilibrates oxygen dissociation (Eq.~\ref{eqn:Odiss_Kp}) but forces CO product gas $(\Delta_{\mathrm{C:O}}=0.75$).
The roots of Eq.~\ref{eqn:interfacecoupling} for free oxygen at the wall are solved. 
Table generation follows the procedure proposed in the prior section, beginning with $M_w=\SI{28.97}{g/mol}$ and iterating until convergence. 

\begin{figure}[tb!]
\centering
\begin{subfigure}{\columnwidth}
    \centering
   \includegraphics[width=.35\paperwidth]{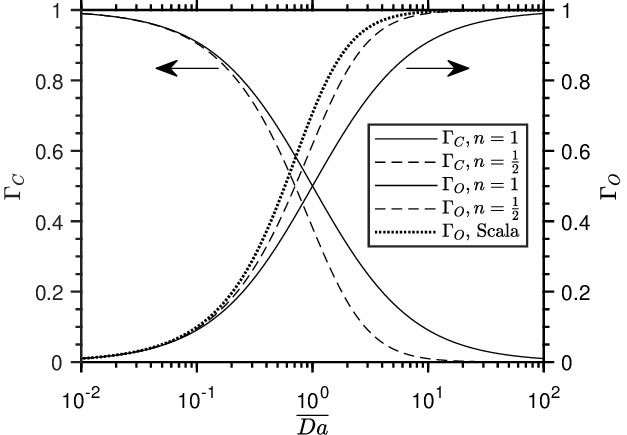}
   \caption{$\Gamma_O$ and $\Gamma_C$.}
   \label{fig:Delta}
\end{subfigure}
\begin{subfigure}{\columnwidth}
    \centering
   \includegraphics[width=.35\paperwidth]{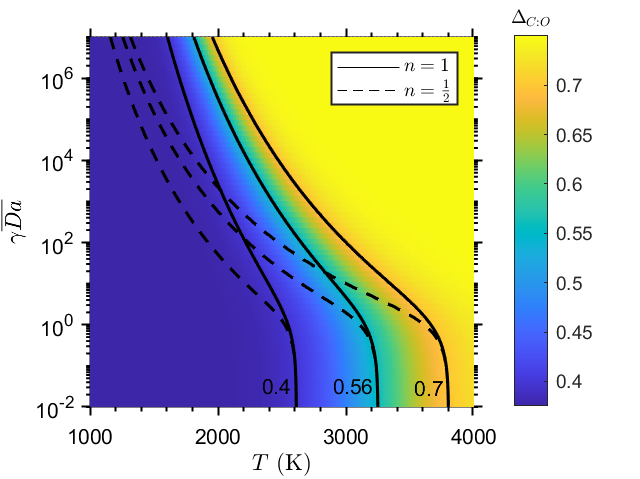}
   \caption{Equilibrium $\Delta_{\mathrm{C:O}}$ for $P=\SI{1}{atm}$.}
   \label{fig:Delta}
\end{subfigure}
\begin{subfigure}{\columnwidth}
    \centering
   \includegraphics[width=.35\paperwidth]{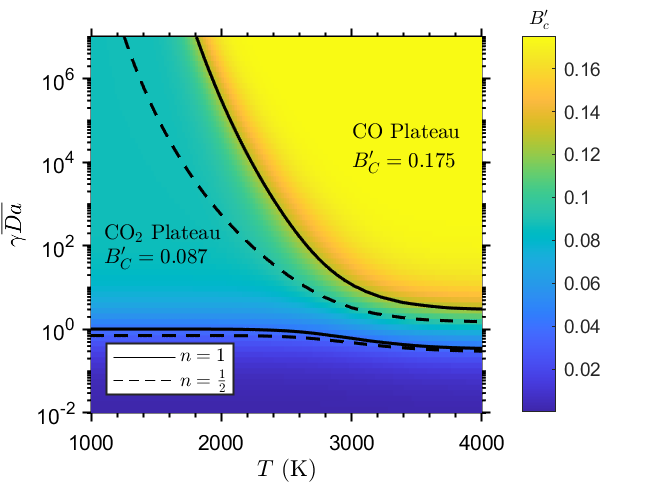}
   \caption{Equilibrium $B'_C$ at $P=\SI{1}{atm}$ for $n=1$ with $n=\frac{1}{2}$ contours overlaid.}
   \label{fig:Delta}
\end{subfigure}
\caption{Damk\"ohler Thermochemistry Tables for $B'_{c,s}=0$.}
\label{fig:Da_tables}
\end{figure}

Fig.~\ref{fig:Da_tables} presents the results of the Damk\"ohler thermochemistry tables for $B'_{C,s}=0$ and equilibrium wall-gas chemistry. 
Fig.~\ref{fig:Da_tables}a compares the $\Gamma$ functions to the half-order approximation proposed by Scala \cite{Scala1965}.
Half-order reactions approach the diffusion limit at lower Damk\"ohler number by roughly an order-of-magnitude. 

In Fig.~\ref{fig:Da_tables}b, $\Delta_{\mathrm{C:O}}$ is bounded by 0.375 (CO$_2$) and 0.75 (CO). 
The product gas is predominantly independent of $\overline{Da}$ and $n$ in the kinetically limited regime ($\overline{Da}\ll1$).
Once diffusion limited ($\overline{Da}\gg1$), higher Damk\"ohler number favors CO formation and $n=\frac{1}{2}$ strengthens this effect.

In Fig.~\ref{fig:Da_tables}c, the influence of the $\Gamma_O$ and  $\Delta_{\mathrm{C:O}}$ partitions ablation into three distinct regimes: kinetically limited ablation for $Da\ll1$, and diffusion-limited behavior in CO$_2$ ($B'_c=0.087$) and CO ($B'_c=0.175$) plateaus for $Da\gg1$.  
 For non-equilibrium CO tables, $\Delta_{\mathrm{C:O}}=0.75$ and $B'_C$ is dependent solely on $\gamma \overline{Da}$ (Fig.~\ref{fig:Da_tables}a).

$B'_{C,s}>0$ is not visualized here, but primarily raises the floor in the kinetically limited regime with muted effects otherwise.
Notably, the limit of Eq.~\ref{eqn:Delta_Kp} as free oxygen approaches zero is $\Delta_{\mathrm{C:O}}=0.75$.
Then by definition, $B'_{C,s}>0.087$ elicits CO production enforcing $\Delta_{\mathrm{C:O}}> B'_{C,s}/\tilde{y}_{\mathrm{O},e}$ 
and the upper limit of the oxygen-rich tables is always $B'_{C,s}=0.75 \tilde{y}_{\mathrm{O},e}$ under equilibrium.

\subsection{Table Lookup and Predictions}

To demonstrate table output for ablation cases, a first-order reaction set deploys the oxidation model of Park \cite{Park1976} and the carbon sublimation model from Zhluktov and Abe \cite{Zhluktov1999,Zavitsanos1966}, each reaction taking the form:
\begin{equation}
    k_{j}= \varepsilon_j \sqrt{\tfrac{1}{2\pi} \tfrac{R}{M_j} T}
\end{equation}
for species O, O$_2$, C$_1$, C$_2$, and C$_3$.
For half-order kinetics, the oxidation model is instead Scala's `slow' model \cite{Scala1962}, which neglects dissociation and is here applied to the averaged oxygen consumption kinetics ($\bar{k}_{\mathrm{O}_x}$). 

\begin{figure}[tb!]
\centering
\begin{subfigure}{\columnwidth}
    \centering
   \includegraphics[width=.35\paperwidth]{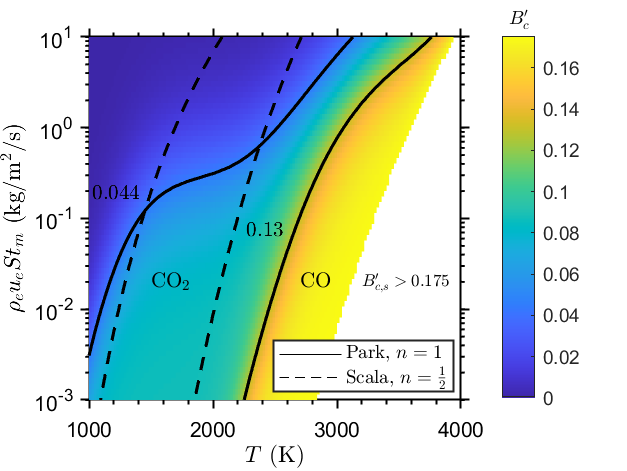}
   \caption{Equilibrium Wall Gas, Park model overlaid with Scala contours.}
   \label{fig:kinetics_equil}
\end{subfigure}
\begin{subfigure}{\columnwidth}
    \centering
   \includegraphics[width=.35\paperwidth]{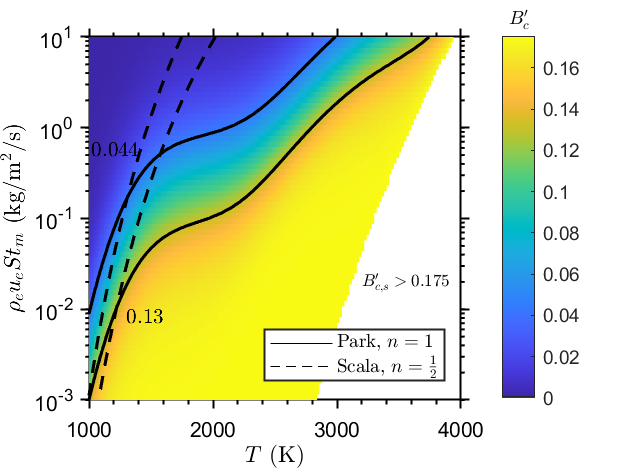}
   \caption{Non-equilibrium Wall Gas, Park model overlaid with Scala contours.}
   \label{fig:kinetics_CO}
\end{subfigure}
\begin{subfigure}{\columnwidth}
    \centering
   \includegraphics[width=.35\paperwidth]{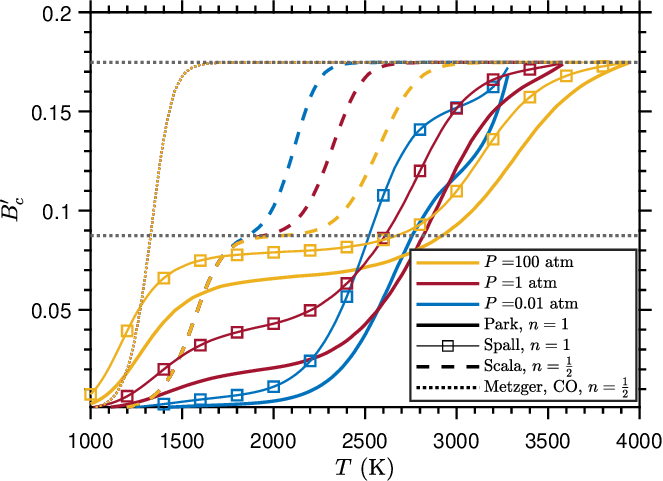}
   \caption{Ablation predictions at a stagnation point.}
   \label{fig:kinetics_spall}
\end{subfigure}
\caption{ Ablation-rate predictions for surface-kinetics models.}
\label{fig:kinetics_tables}
\end{figure}

The table lookup procedure follows: 
\begin{enumerate}
    \item Estimate $y_{\mathrm{O},diss}$ \& $\rho_w$.
    \item Calculate $\gamma\overline{Da}$ and $B'_{C,s}$.
    \item Table lookup for new wall-gas state. 
    \item Converge Steps 1-3.
    \item Table lookup of converged values. 
\end{enumerate}
Fig.~\ref{fig:kinetics_tables} presents results assuming $\gamma=1$ using the thermochemistry tables from Sec.~\ref{sec:tablegen}. 
Fig.~\ref{fig:kinetics_tables}(a) reveals distinct CO and CO$_2$ plateaus under homogeneous equilibrium.
Consistent with literature \cite{Scala1965,Welsh1963}, the half-order model appears more reactive, especially at high mass conductance.
These oxygen-rich tables end with the limit set by Eq.~\ref{eqn:Bprime_regimes}. 
Fig.~\ref{fig:kinetics_tables}(b) presents the non-equilibrium result for CO ablation product, which enhances ablation rate for both models.

Applying these results to an ablation case, consider a stagnation point with radius $r=\SI{1}{cm}$.
Mass conductance is calculated from stagnation pressure using Scala's correlation for supersonic stagnation points \cite{Scala1962}, where $g_m \propto \sqrt{P/r}$.
Fig.~\ref{fig:kinetics_tables}(c) presents the predicted ablation rates across temperature and pressure.
The first-order gamma function approaches unity faster at higher pressure. 
The half-order gamma functions are identical across pressure conditions, yielding equivalent ablation rate below the CO$_2$ plateau.
This unique result is traceable to the Damk\"ohler number (Eq.~\ref{eqn:Damkohler}), where $\sqrt{\rho_w}$ and $\rho_e u_e \mathit{St}_m$ are both proportional to $\sqrt{P}$. 
In both cases, the transition to CO is delayed by increased pressure.
Welsh \cite{Welsh1963} reported similar trends but with pronounced convergence to the CO$_2$ plateau for extended temperature spans.

Applying the half-order reaction proposed by Metzger \cite{Metzger1967}, the Damk\"ohler thermochemistry model for CO ablation product is in close agreement with the published results. 
Variance is attributable to the choice of $\Gamma_\mathrm{O}$ and the molecular-weight ratio in Eqn.~\ref{eqn:Da_compare}.
As observed by Metzger, the ablation rate for this model is entirely independent of pressure.

Finally, the first-order model is augmented with the spallation model of McVey \cite{McVey1970} (Model A), combusting all spall at the wall via $B'_{\mathrm{C},s}$.
This model increases ablation rate by an analytical derivation representing the critical height at which a graphite filler particle separates due to shear and accelerated oxidation of the binder phase:
\begin{equation}
    B'_{C,s}=B'_{C,oxy} \frac{ \left(\tfrac{h}{a}\right)_{cr} +\tfrac{\rho_f}{\rho_b} -1}{ \left(\tfrac{h}{a}\right)_{cr}+\tfrac{\rho_f}{\rho_t}-1}
\end{equation}
Using McVey's recommended values for densities of the filler ($\rho_f=\SI{2.15}{g/cm^3}$), binder ($\rho_b=\SI{1.2}{g/cm^3}$), and bulk ($\rho_t=\SI{1.8}{g/cm^3}$) and arbitrarily setting the critical height ratio to $\left(\tfrac{h}{a}\right)_{cr}=0.4$, the total ablation rate is roughly doubled relative to surface oxidation, but the resulting spall consumes free oxygen, diminishing the overall impact as Damk\"ohler number approaches unity. 
An engineering model would typically predict the critical height based on shear \cite{Ziering1972,Gosse2009}, but a detailed consideration of spallation physics is beyond the present scope. 

%% file: conclusion.tex
\section{Discussion}
\label{sec:discussion}

The Damk\"ohler thermochemistry tables provided excellent flexibility for testing various model forms and kinetics parameters. 
Using a set of four tables (equilibrium and CO ablation product for first and half-order reactions), the kinetics-model demonstration in Figure~\ref{fig:kinetics_spall} relied on a simple table lookup procedure (Sec.~\ref{sec:demonstration}.2). 
This capability for testing and varying kinetics-models appears the principle advantage of the present approach and may prove compatible with various empirical and computational objectives.

Returning to the literature that inspired this work, the Damk\"ohler number provides a more formal definition for determining kinetically limited and diffusion-limited conditions, as well as the transition therebetween.
The present results buttress the ablation-ratio methodology: the wall-gas molecular weight was typically bounded between approximately \SI{28}{g/mol}  (CO plateau) and \SI{31.4}{g/mol} (CO$_2$ plateau). Consequently, wall-gas molecular weight does not vary greatly from the reference state (\SI{28.97}{g/mol}, Eq.~\ref{eqn:Da_compare}), and the ablation-rate ratio remains a simple and effective approximation of the Damk\"ohler number in air-carbon systems. 

The single- and double-plateau predictions arising from ablation product assumptions (equilibrium vs. CO) appears a useful engineering tool, but lacks predictive power. 
Specifically, low-pressure ($<$\SI{1}{atm}) ablation data \cite{Metzger1967} suggests immediate convergence with the CO plateau, but the half-order, equilibrium-product predictions in Fig~\ref{fig:kinetics_spall} are remarkably consistent with intermediate ablation rate ($0.087<B'_c<0.175$) in Perini's compilation of high-pressure cases \cite{Perini1971}. 
An alternate method for calculating $\Delta_{C:O}$ with approximations of finite-rate gas-phase chemistry could plausibly predict these pressure-dependent observations.

Similarly, the thermochemistry tables force a particular definition for the reaction order.
Although numerical solutions to Eq.~\ref{eqn:interfacecoupling} for any predefined reaction order are likely obtainable, a solution procedure for advanced, surface-coverage dependent models \cite{Prata2021} remains unclear. 

The newly defined $\gamma \overline{Da}$ term drives ablation throughout the majority of the thermochemistry table. The direct-carbon source ($B'_{\mathrm{C},s}$) from sublimation alone is typically negligible, but can drive ablation rate to the CO plateau at high temperature 
and accommodate engineering models for spallation.
In many cases, this variable could be removed from the table lookup procedure -- collapsing the table to three dimensions ($T$,$P$,$\gamma\overline{Da}$).

Iterative thermochemistry solutions resolved coupling between the kinetics model and the wall-gas state in the present work, but may prove infeasible for physics coupling in modern ablation codes.
Extensibility beyond simple air-carbon systems also remains unclear. 
Further development may be necessary to reformulate the solver procedure or tabulation variables. 
The current implementation is therefore limited to empirical analysis of kinetics models and ablation data. 
Established techniques may remain preferred for decomposing ablators, multimaterial systems, and high-fidelity models \cite{Milos1997}.

\section{Conclusion}
\label{sec:conclusion}

The presented derivation enables ablation-rate predictions for finite-rate surface kinetics outside of a traditional ablation thermochemistry routine. 
Various kinetics models for oxidation, sublimation, and spallation were demonstrated, quantified and tabulated via newly defined non-dimensional kinetics terms. 
Future development could pursue integration into high-fidelity simulations, 
improved predictions of homogeneous non-equilibrium, and empirical analysis of finite-rate ablation datasets.